# On the interplay of paramagnetism and topology in the Fe-based High $T_c$ Superconductors

J.D. Rameau, N. Zaki, G.D. Gu, P.D. Johnson

*Condensed Matter Physics and Materials Science Division, Brookhaven National Laboratory, Upton, New York 11973*

M. Weinert

*Department of Physics, University of Wisconsin-Milwaukee, Milwaukee, Wisconsin 53201*

The high $T_c$ superconductor, FeTe$_{0.55}$Se$_{0.45}$ has recently been shown to support a surface state with topological character. Here we use low-energy laser-based ARPES with variable light polarization, including both linear and circular polarization, to re-examine the same material and the related FeTe$_{0.7}$Se$_{0.3}$, with larger Te concentration. In both cases we observe the presence of a surface state displaying linear dispersion in a cone-like configuration. The use of circular polarization confirms the presence of helical spin structure. These experimental studies are compared with theoretical studies that account for the local magnetic effects related to the paramagnetism observed in this system in the normal state. In contrast to previous studies we find that including the magnetic contributions is necessary to bring the chemical potential of the calculated electronic band structure naturally into alignment with the experimental observations.

The two key research areas of high $T_c$ superconductivity and topological insulators have converged in studies of $FeTe_{0.55}Se_{0.45}$ where it has recently been reported that this high $T_c$ superconductor supports a topological surface state at the center of the Brillouin zone.[1] Furthermore recent studies indicate the possible existence of Majorana Fermions or zero energy modes in the superconducting state.[2] Majorana Fermions (MF), long sought-after in high energy particle physics, represent particles that are their own anti-particles. Aside from the fundamental interest in such exotic particles, they may have application as especially robust components for quantum computing. As such there has been a considerable effort to identify MF in different heterostructures involving the interface of a topological insulator with a superconductor. The original proposal was for proximity to an s-wave superconductor[3] although experiments have also been attempted with unconventional d-wave superconductors where the defining superconducting gap is larger.[4,5] The observation of MF has also recently been made in studies of a quantum anomalous Hall insulator grown in contact with an s-wave superconductor.[6] However the latter phenomenon was only observed at T~1°K and in very fragile device structures. It is therefore of considerable interest if there exists a high $T_c$ superconductor that supports an inherent topological state that is susceptible to the superconducting transition. While the discussion here focuses on the family $FeTe_{1-x}Se_x$ (FTS), the potential for the same phenomena to exist in other Fe-based superconductors clearly exists. Indeed a recent report suggests that this is indeed the case.[7]

Topological insulators represent materials that are insulating in the bulk but metallic in the surface layer. In non-magnetic systems with inversion symmetry, it is possible to characterize insulators by a $Z_2$ topological index, obtained by taking the product of the parities at the time-reversal-invariant points in the occupied Brillouin zone.[8] A value of +1 for $Z_2$ describes a conventional insulator, a value of -1 describes a nontrivial topological insulator. Recognizing that the atomic structure of the FTS system is characterized by the inversion symmetric crystallographic space group P4/nmm, previous theoretical studies of $FeTe_{0.55}Se_{0.45}$ have applied the $Z_2$ analysis and concluded that the material might indeed support a topologically nontrivial band structure under certain assumptions.[9] Such a conclusion does appear to be supported with the recent experimental observation of Dirac-like Fermions at the center of the Brillouin zone in this material.[1] However, calculated non-magnetic (NM) band structures place the relevant bands and associated spin-orbit induced inverted gaps well above the chemical potential.[9,10] Attributing such shifts of the chemical potential to doping[9] are not a viable solution since the bands around the $\bar{M}$ point will also be shifted and consequently pushed into significant disagreement with experiments. We note that the requirement that the band gap supporting the surface state is in alignment with the chemical potential as observed experimentally has also lead to the discussion of the role of vertex interactions and band renormalizations leading to a "curved chemical potential".[10] In the present study we revisit these systems both experimentally and theoretically focusing on the normal state. We are able to confirm and extend the previous experimental observations. However theoretically we show the importance of recognizing the paramagnetic – as opposed to non-magnetic – character of the normal state.

The $FeTe_{1-x}Se_x$ family has a complex phase diagram. FeTe itself is a non-superconducting material. Doping with Se leads to the possibility of superconductivity. In fact the complete replacement of Te by Se results in FeSe, a superconductor. Grown at the monolayer level, FeSe displays the highest superconducting transition temperature found in the Fe-based high $T_c$ materials. In our own previous studies of the $FeTe_{0.55}Se_{0.45}$ system, we pointed to the role of spin-orbit coupling in lifting the degeneracy of the bands at the center of the zone and its potential role in the observed nematicity in these systems.[11] The associated photoemission studies were carried out at higher photon energies with lower energy and momentum resolution. While they confirmed the role of spin-orbit interactions in lifting the degeneracy, they did not provide clear evidence of any topological state, although in hindsight the latter state was already evident in second-derivative plots shown in that study. Here we study the same material again but now using high-resolution laser based ARPES at 6 eV photon energy with variable light polarization, including both linear and circular polarized incident light. With the higher energy and momentum resolution we are able to confirm that both $FeTe_{0.55}Se_{0.45}$ and the related $FeTe_{0.7}Se_{0.3}$ do indeed support electronic states in the form of Dirac cones with a helical spin structure as expected for a topological surface state. These experimental studies are compared with theoretical studies within the framework of DFT that, unlike the previous studies,[9,10] take account of magnetic effects intrinsic to the paramagnetism observed in the normal state.[12] As noted previously, it is often necessary to take account of such magnetic contributions to accurately describe the physical properties of the Fe-based superconductors.[13,14] However it is important to also note that the paramagnetism characterized by the (disordered) local moments can potentially remove the time reversal symmetry required in the normal analysis of topological insulators and will significantly alter the electronic states. Further such breaking of time reversal invariance leads to the possibility of a gap associated with mass acquisition at the Dirac point as has been observed previously in the case of a magnetically doped topological insulator.[15] As shown below, including the magnetic contributions removes the necessity of arbitrarily adjusting the chemical potential as the calculated electronic band structure moves naturally into alignment with the experimental observations. With this realignment, the discussion of the nature of the band gap supporting the topological surface state does indeed change. Finally we note that the FTS material is not an insulator, being more metallic in character and thus its characteristics are probably more akin to those of a topological metal or semimetal.

The experimental studies reported here were carried out using a 3 psec pulse width, 76 MHz rep rate Coherent Mira 900P Ti:Sapphire laser, the output of which was quadrupled to provide 6.0 eV incident light, focused into a spot on the sample ~ 20 μm in diameter. The polarization of the latter could be varied with the use of quarter and/or half wave plates to provide linear or circularly polarized light of arbitrary orientation on the Poincare sphere. Photoemission spectra were obtained using a Scienta SES 2002 electron spectrometer. Single crystals of $FeTe_{1-x}Se_x$ were grown by a unidirectional solidification method. The nominal compositions had no excess Fe, and $T_c$, measured by magnetic susceptibility, is 14 K in both samples. These samples were cleaved *in situ* at T≤ 20 K and base pressure of $5x10^{-11}$ Torr with measurements on each sample completed within 8 hours. The surface quality of each measurement location was confirmed by LEED at the end of each measurement and the $E_F$ referenced to a gold sample in contact with the FTS samples. The effective energy resolution is ~4meV (FWHM). The angular resolution was ~ $0.002Å^{-1}$ at the low photon energy used.

We firstly examine these materials comparing different doping levels in the same family. In figure 1 (a) we show the spectral response obtained from an $FeTe_{0.55}Se_{0.45}$ system using linearly polarized light, in this case p-polarized corresponding to the polarization vector being more perpendicular to the surface. In Figs. 1(b) and (c) we show the spectral response obtained from the related $FeTe_{0.7}Se_{0.3}$ system again using linearly polarized light, but now with p and s polarized light, the latter corresponding to the polarization vector being in the surface plane. In the lower panels, (d), (e) and (f) we show the energy distribution curves EDCs derived from the upper panels for different $k_\parallel$. The presence of the Dirac cone is clearly visible in all panels. However the cone is more visible with p-polarized light than s-polarized light. Interestingly the spectral response in panel (a) also appears to highlight flatter bulk bands, which we associate with the bulk Z-point ($k_\parallel=0$, $k_z=\frac{1}{2}$).

In Figs. 2 (a) and (b) we show the spectral response that we obtain from the same region of the Brillouin zone with the same incident photon energy but now using circularly polarized light. The two panels show the effect of switching from left- to right-hand circular (LHC and RHC, respectively) polarization. The Dirac state clearly shows the appropriate sensitivity to light polarization characteristic of a helical spin structure, the latter an indication of the surface nature of the state. We show this further in figure 2(c) where we present momentum distribution curves MDCs showing the dichroism as a function of binding energy. The dichroic sensitivity switches either side of the Dirac point, ~ 8.0 meV below the chemical potential. However we note that the measured dichroism of 30% from the FTS sample as defined by $(I_L - I_R)/(I_L + I_R)$ is smaller than that measured under the same experimental geometry from the Dirac state on the classic topological insulator $Bi_{1.5}Sb_{0.5}Te_{1.7}S_{1.3}$, which is of the order of 80% in our own studies.[16] The difference may reflect a different degree of spin texture in the two systems or a greater overlap between the surface state and the bulk bands in the more metallic FTS system. We further note that the observation of dichroism when using incident circularly polarized light can also reflect polarization effects in the final state. However we would suggest that such an observation represents an unlikely explanation in view of the switch in measured dichroism within a few meV either side of the Dirac point.

We now examine how these experimental observations compare with theoretical calculations. Experimentally the FTS (and related FeSe) materials do not have long-range magnetic order, but the Fe atoms have strong Hund's rule local magnetic and nearest neighbor antiferromagnetic correlations. Capturing the physics of this disordered magnetic state is not a simple task, and typically the electronic structure is calculated in ordered magnetic configurations or, as noted above, by assuming that the paramagnetic state (with average zero moment) is equivalent to zero magnetization everywhere. However as described elsewhere,[17] the paramagnetic state can be modelled as an explicit configurational average of a (large) number of different "snap shots" to mimic the various short-range interactions and configurations that might occur. Experimentally there is good overall agreement with the bands calculated for the so-called checkerboard antiferromagnetic (CB-AFM) ordering despite the fact that this is not the lowest energy ordered magnetic configuration. Again as discussed in more detail elsewhere,[17] DFT spin-spiral calculations place this $S=1$ spin system in a region of phase space where CB-antiferromagnetic (AFM) quantum fluctuations necessarily lead to a magnetically disordered paramagnetic phase whose electronic structure will closely resemble that of the ordered $q=0$ CB configuration. (For the DFT calculations to be consistent with the observed paramagnetic phase, the calculated mean-field groundstate must be the collinear stripe phase as found.) Thus, we model the FeTe$_{0.5}$Se$_{0.5}$ system using a c(2x2) Te-Se ordering (above and below the Fe plane) and with the Fe moments in the CB AFM ordering, which has the same lowest order radial correlation function as the random FeTe$_{0.5}$Se$_{0.5}$ alloy. Both magnetic and Te-Se compositional fluctuations will lead to M-point contributions to the electronic structure. . All the resulting bands are then $k$-projected back to the crystallographic cell. Since the topological protection is a bulk property, we consider bulk calculations only. The latter do not show evidence of a topological surface state (TSS) directly. However we note that DFT calculations of the same system published elsewhere also did not show direct evidence of a TSS. Rather the latter calculations relied on an effective Hamiltonian approach superimposed on the DFT calculations to identify potential TSS. Our calculations are performed using the Full-potential Linearized Augmented Plane Wave method, including spin-orbit, with a plane wave cutoff of 275 eV for the wave functions. Because of the different atomic sizes of Se and Te the heights above the Fe planes differ by ~ 0.3 Å, which were then fixed for all calculations.

In Fig. 3 we compare the calculated band structures in the Γ – Z direction for non-magnetic vs. magnetic, and with/without the inclusion of spin-orbit coupling (SOC). The orbital character of the states at the zone boundaries are shown for the case of no SOC. We can make a number of observations immediately. In the absence of local magnetic moments, the relevant calculated bands, both with and without SOC, are well above the chemical potential, in agreement with previous calculations.[9,10] The earlier non-magnetic calculations point to the possibility of a gap (indicated by the circle in Fig. 1a, lower panel) capable of hosting a topological state arising from the SOC-induced hybridization of the downward dispersing band composed of Fe d$_{x2-y2}$ (antibonding between Fe sites) orbitals with the chalcogen p$_z$ orbitals. In the presence of local magnetic moments and the resulting local exchange splitting of Fe orbitals, both the flat $d_{x2-y2}$ band and the downward dispersing $p_z$ band are pushed essentially out of the region of interest. In addition, as indicated in Fig.3b (upper left panel) the character of the two bands interchange between Γ and Z, indicating the type of band inversions conducive to the formation of topological states. The effect of SOC on the bands for the magnetic case, Fig.3b, differs depending on the assumed easy direction of the moments. For in-plane moments (pointing along the *x* direction), the separation between the bands increases and the upper band at Z crosses the Fermi level. (Using a *k*-projection analysis, the bands seen in thin film calculations near the Fermi level around Γ can be identified as coming from bulk states around Z.) For perpendicular moments (*z* direction), the resulting SOC bands are more complicated. Because of the broken time-reversal symmetry, the states at *k* and -*k* are no longer degenerate, although still periodic, and the wave function character shifts among the different branches as a function of $k_z$, with various gaps opening. The maximum of the bands is now not at Z but is shifted slightly away. Although the inclusion of magnetism breaks both time-reversal and inversion symmetry, the product remains a symmetry under AFM ordering, with the result that the SOC bands are all doubly degenerate and the two wave functions are related by this product, i.e., inversion transforms orbitals on one Fe site to orbitals on the other Fe site and time-reversal flips spin. Thus, in contrast to the non-magnetic (and "standard" topological insulator) situation, the wave functions cannot be classified by parity at the time reversal invariant momentum (TRIM) points.

In Fig. 4 we show the calculated band dispersion as a function of $k_∥$ for both in-plane and out-of-plane moments at a $k_z$ corresponding to the maximum of the band displacement (e.g., at the *Z* point for in-plane moments) for several different directions. As a result of magnetic ordering and static compositional fluctuations, the calculated bands for both in-plane and perpendicular moments show behavior indicative of the type of inverted gaps capable of supporting a topological state: the upper band has its maximum – and largest spectral weight – around $k_∥=0$, and disperses downward until it reaches a minimum at the gap, at which point the spectral weight and wave function character (determined by tracking the overlap of bands as a function of *k*) shift to the bands on the other side of the gap. As expected, the wave function character of the upper band for $k_∥$ further out and the lower band on the other side of the gap near $k_∥=0$ show the complementary behavior. As seen in fig. 4(d) the gap exists for a large range of $k_z$ and over the whole $k_∥$ region of interest. Moreover, note the saddle-shaped topology characteristic of band inversion giving two maxima in the lower sheet. Because of the mixing of spatial and magnetic degrees of freedom in the symmetry operations, the inversion of the gaps involves both orbital and spin character. As discussed above, these band features are expected to be robust in the paramagnetic phase, and thus the latter phase may indeed host a topological Dirac surface state in the symmetry gaps around Z of the bulk. While such states are not evident in bulk calculations finite film calculations of up to 13 FTS layers do not show evidence of a topological Dirac state either, likely indicating that the relevant bulk band features are not developed well enough in these thin films to support a TSS. As a crude estimate of the size of the film required to support a TSS we note that to describe the dispersion (curvature) of the bulk bands around the extrema of the bands by a series of discrete points requires a spacing $\Delta k_z$ less ~10% of Γ-Z, which corresponds to > 20 layer film. (For prototypical $Bi_2Se_3$, the minimum thickness is 5-6 QLs.)

As noted earlier, the symmetry breaking associated with the local moments, evident in the lower panel of fig. 3(b) and fig. 4(b), leads to the possibility of the topological state acquiring mass via the Haldane mechanism (a magnetic field that is zero on the average but has the full symmetry on the lattice),[18] not unlike the paramagnetic state discussed here. Further, it has been shown that doping with magnetic Fe atoms leads to mass acquisition.[15] In fact in the latter study the small gaps observed at low doping were believed associated with a paramagnetic state. However it has also been suggested that the Dirac electrons on the surface can interact through an RKKY interaction resulting in a ferromagnetic surface[19] when the chemical potential lies near the Dirac point, precisely the situation existing in the present systems. That the possibility for a gap and hence mass acquisition exists is evident in Figs. 1. Clearly in the latter figure the dispersion of the states in the immediate vicinity of the Dirac point do not look linear but rather more "quadratic". We believe that such a gap with associated dispersion is also a characteristic of the earlier studies of the normal state published elsewhere.[1] The possibility is also evident in the spectra of Fig. 2 which again appear to indicate the presence of a gap in the vicinity of the Dirac point. However a word of caution is required. Examination of the proposed Dirac cone for this system indicates that a minimal misalignment ($\leq 0.5^0$) could give the appearance of such a gap near the Dirac point. Resolution of the latter issue will ultimately require a more precise measurement.

In summary we have confirmed that a surface state with all the characteristics of a topological state does indeed exist on the surface of members of the Fe-based superconductor family FeTe$_{1-x}$Se$_x$. Furthermore our calculations suggest that the state exists in a spin-orbit induced gap near the Z-point in the bulk Brillouin zone. The gap has the required band inversion to support a topological state. We note that we have earlier reported detailed investigations of the bulk bands at the Z-point involving different photon energies.[11] However the observation of the topological state will occur with a wide range of photon energies reflecting its localized nature in the surface layers.[20] We also make the important observation that correct alignment of the calculated bands with experimental observation as found in the present study reflects the recognition that the normal state is characterized by the disordered local moments associated with the paramagnetic ground state. The latter potentially breaks the time reversal and inversion symmetries suggesting the topological protection is more akin to that found in topological metals such as Weyl or Dirac semi-metals.

The authors acknowledge useful discussions with Igor Zalyznyak and John Tranquada, The experimental work carried out at Brookhaven was supported by the U.S. DOE under Contract No. DE-AC02-98CH10886 and in part by the Center for Computational Design of Functional Strongly Correlated Materials and Theoretical Spectroscopy. The theoretical studies (MW) at UWM were supported by the Department of Energy (DE-SC0017632).

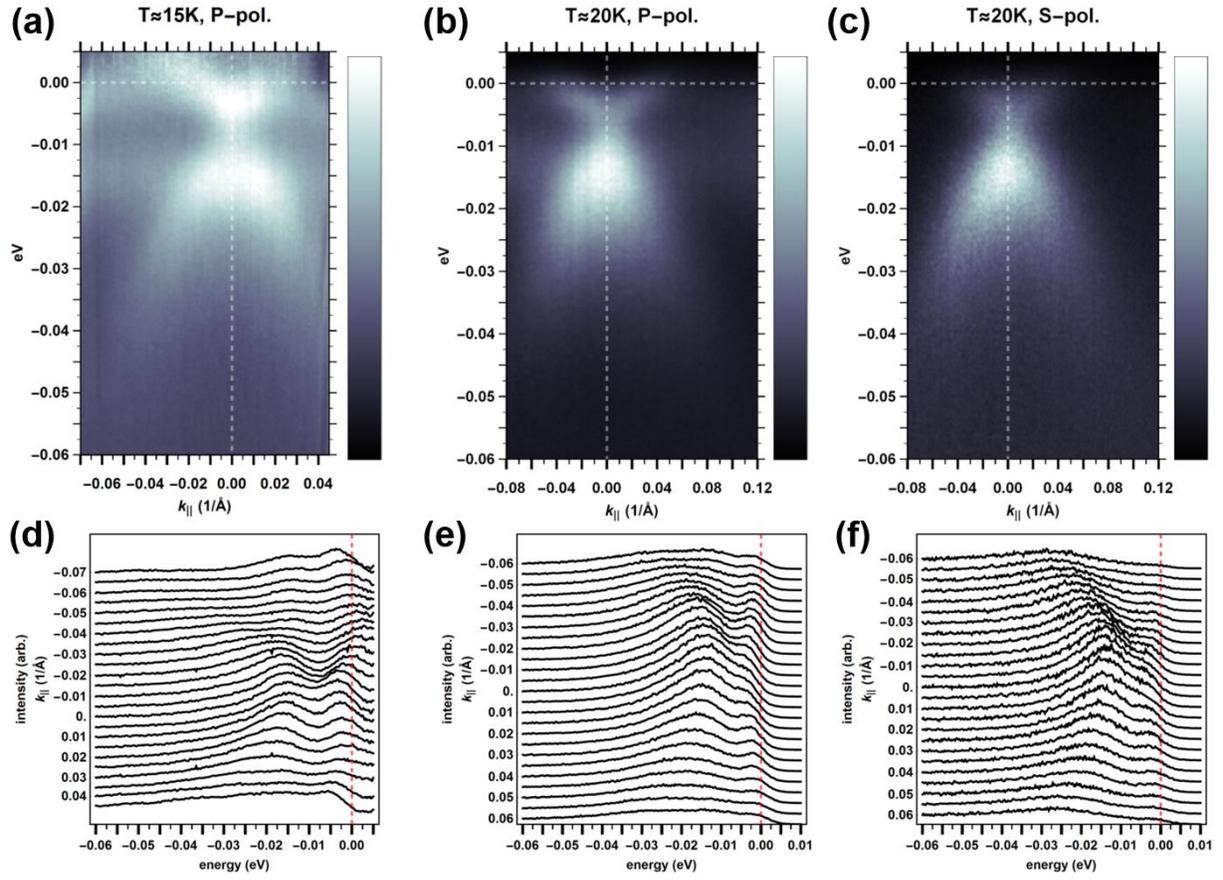

Fig. 1. a) ARPES measurement of the electronic structure around the $\Gamma$ point of FeTe$_{0.55}$Se$_{0.45}$ using p-polarized light, b) same measurement as in a) but now FeTe$_{0.7}$Se$_{0.3}$ using p-polarized light, c) same measurement as in b) but using s-polarized light. d) – f) Cuts through the spectral maps in a) – c) showing EDCs in the vicinity of the $\Gamma$ point. The temperatures associated with the different measurements are indicated in a) – c).

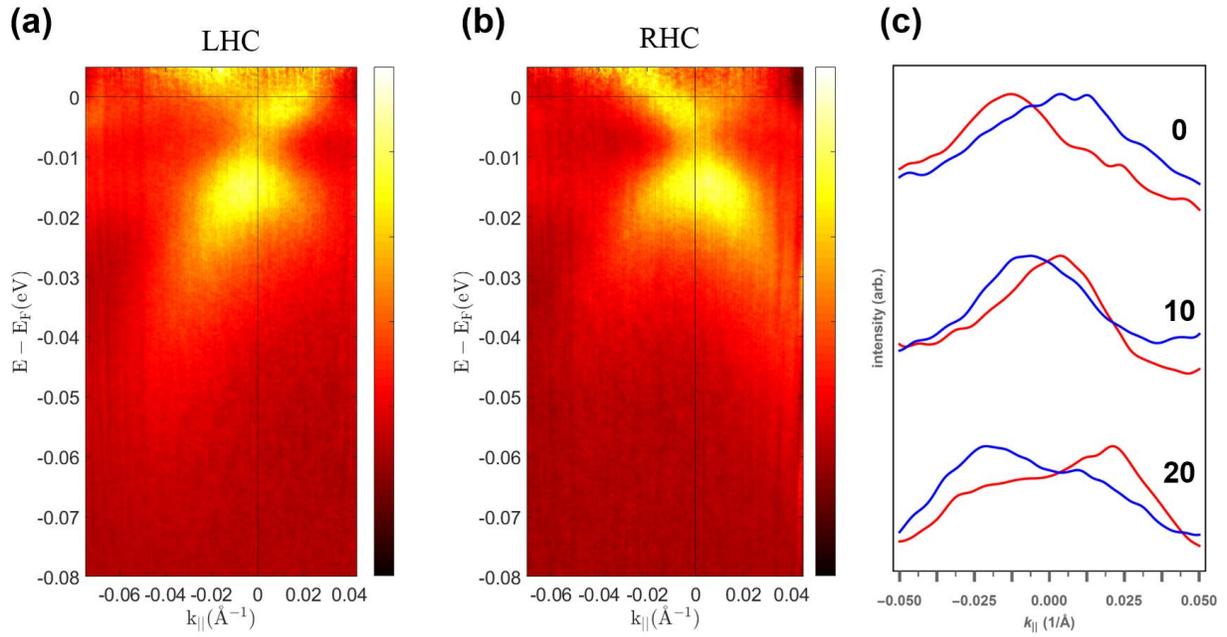

Fig. 2. The Dirac cone as seen on FeTe$_{0.55}$Se$_{0.45}$ using a) left-circularly polarized light and b) right-circularly polarized light. The Dirac point is located at approximately 8.5 meV below the chemical potential   In c) we show MDC cuts through spectra measured with left polarized (red) and right polarized (blue) light as a function of binding energy (meV) as indicated.  In all cases the sample is held at 15K in the normal state.

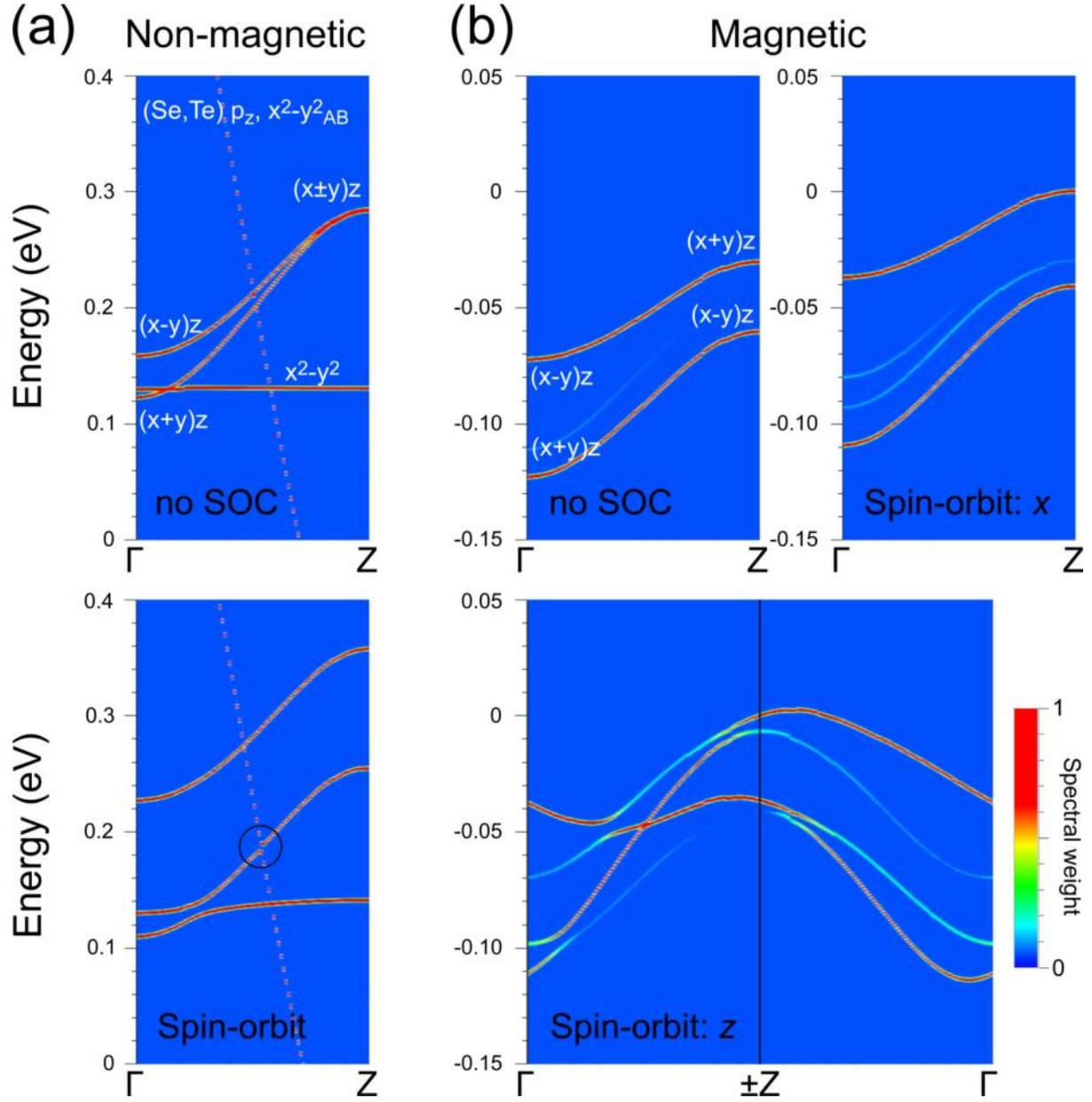

Fig. 3. Calculated *k*-projected band structures along Γ-Z. (a) Non-magnetic bands with (lower panel) and without (upper) spin-orbit coupling. The SOC-induced gap is indicated by the circle. (b) Magnetic bands without SOC and with SOC for in-plane (*x*) and perpendicular (*z*) moments. Note the different energy scale (factor of 2) between (a) and (b) and the position of the energy zero (set to the calculated Fermi level).

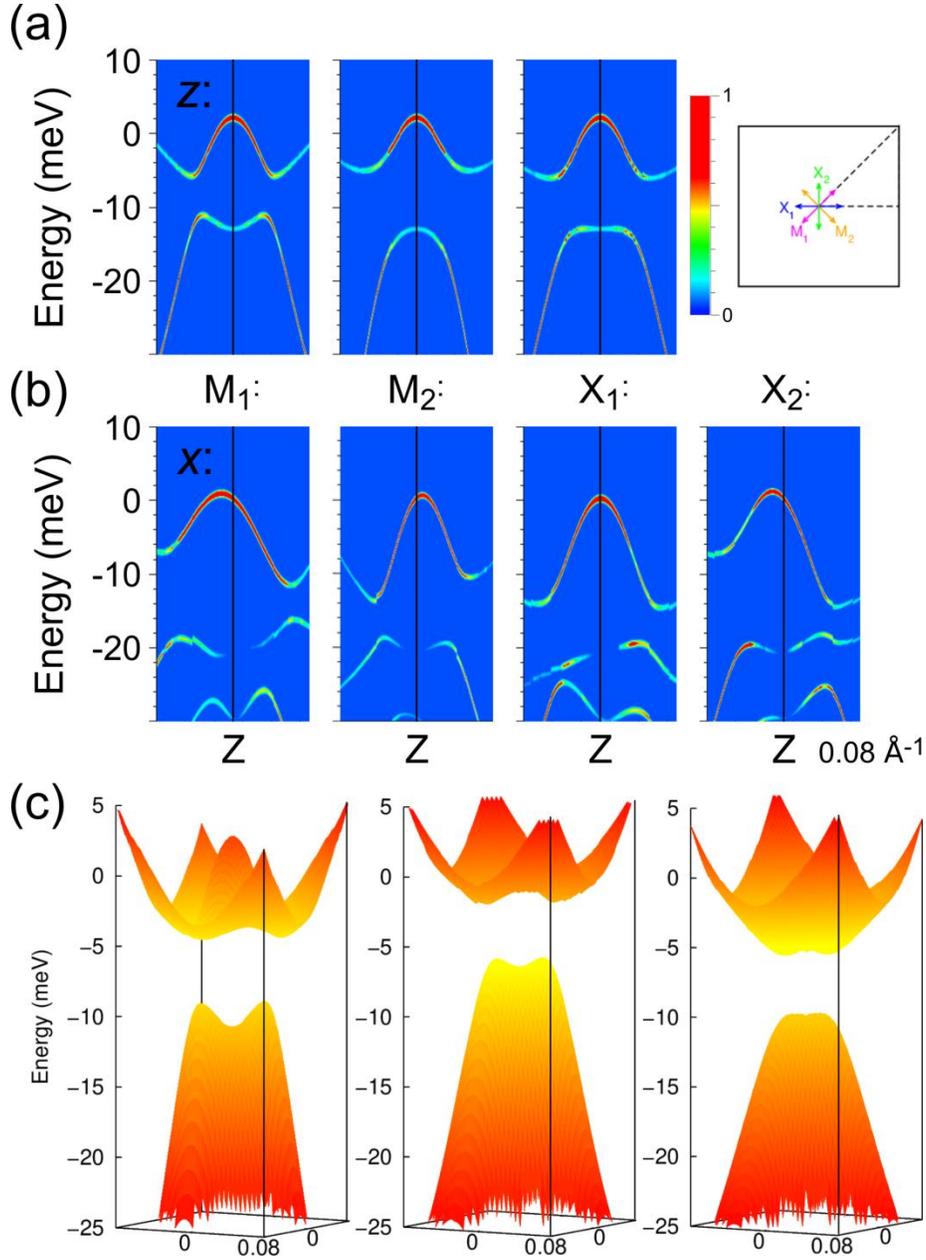

Fig. 4. Calculated k-projected bands for $k_\parallel$ in $k_z$ planes at the band maxima about Z for (a) perpendicular (z) and (b) in-plane (x) moments along different directions as shown in the schematic in the upper right panel. The color corresponds to the calculated spectral weight; since bands of similar character (determined by the overlap with respect to *k*) are found to have similar spectral weight, the spectral weight becomes a proxy for wave function character in the plots. (c) Bands for perpendicular moments over a 0.08 Å$^{-1}$ x 0.08 Å$^{-1}$ region for $k_z$ near the maximum (~0.4 Z), at Z, and at ~0.6 Z. Note the two maxima in the lower band, and that the gap remains for each $k_z$.